\documentclass[
    amsmath,
    amssymb,
    superscriptaddress,
    aps,
    prx,
    reprint,
    floatfix,
]{revtex4-2}

\usepackage{graphicx}
\usepackage{dcolumn}
\usepackage{glossaries}
\usepackage[version=4]{mhchem}
\usepackage[per-mode=symbol]{siunitx}
\usepackage[most]{tcolorbox}
\usepackage{xcolor}
\usepackage{hyperref}
\usepackage{booktabs}

\hypersetup{
    unicode,
    colorlinks=true,
    urlcolor=black,
    linkcolor=black,
    citecolor=black
}

\renewcommand{\vec}[1]{\ensuremath\boldsymbol{#1}}

\newacronym{bec}{BEC}{Born effective charge}
\newacronym{dft}{DFT}{density functional theory}
\newacronym{md}{MD}{molecular dynamics}
\newacronym{mlip}{MLIP}{machine-learned interatomic potential}
\newacronym{mpb}{MPB}{morphotropic phase boundary}
\newacronym{pes}{PES}{potential-energy surface}
\newacronym{qnep}{qNEP}{charge-aware neuroevolution potential}
\newacronym{rmse}{RMSE}{root-mean-square error}

\newcommand{\gpumd}{\textsc{gpumd}}
\newcommand{\vasp}{\textsc{vasp}}
\newcommand{\dynasor}{\textsc{dynasor}}
\newcommand{\phonopy}{\textsc{phonopy}}

\makeatletter
\let\oldtheequation\theequation
\renewcommand\tagform@[1]{\maketag@@@{\ignorespaces#1\unskip\@@italiccorr}}
\renewcommand\theequation{(\oldtheequation)}
\makeatother

\DeclareSIUnit\angstrom{\text{Å}}
\DeclareSIUnit\atom{\text{atom}}
\DeclareSIUnit\bohr{\text{bohr}}

\newcommand{\hmn}[1]{
  \ensuremath{\begingroup\setupHMN #1\endgroup}%
}

\newcommand{\setupHMN}{%
  \doHMN{-}{\HMNoverline}%
  \doHMN{*}{\HMNminverse}%
  \doHMN{i}{\infty}
}

\newcommand{\doHMN}[2]{%
  \begingroup\lccode`~=`#1
  \lowercase{\endgroup\let~}#2%
  \mathcode`#1="8000
}

\newcommand{\HMNminverse}[1]{\frac{#1}{m}}
\newcommand{\HMNoverline}[1]{\mkern1mu\overline{\mkern-1mu#1\mkern-1mu}\mkern1mu}

\begin{document}

\title{Competing Octahedral Tilting and \texorpdfstring{$B$}{B}-Site Off-Centering Govern the Phase Diagram of \texorpdfstring{CsGe$_{1-x}$Sn$_x$Br$_3$}{CsGe(1-x)Sn(x)Br3} Perovskite Alloys}

\newcommand{\addmsu}{Nanomaterials Physics Research Unit, Department of Physics, Faculty of Science, Mahasarakham University, Maha Sarakham 44150, Thailand}
\newcommand{\addchalmers}{Department of Physics and Astronomy, Chalmers University of Technology, SE-412 96, Gothenburg, Sweden}

\author{Apinya Ngoipala}
\email{apinya.n@msu.ac.th}
\affiliation{\addmsu}
\affiliation{\addchalmers}

\author{Erik Fransson}
\affiliation{\addchalmers}

\author{Paul Erhart}
\affiliation{\addchalmers}

\author{Julia Wiktor}
\email{julia.wiktor@chalmers.se}
\affiliation{\addchalmers}

\date{\today}

\begin{abstract}

Lead-free halide perovskites based on Ge and Sn avoid the toxicity of Pb, but the two cations favor different structural distortions.
The stereochemically active Ge $4s^2$ lone pair drives polar $B$-site off-centering, whereas the larger \ce{Sn^{2+}} cation favors octahedral tilting.
Here, we map how these two instabilities compete across the full composition range of \ce{CsGe_{1-$x$}Sn_{$x$}Br3} using large-scale molecular dynamics simulations with a charge-aware machine-learned interatomic potential trained on hybrid-functional data.
At low temperature the alloy exhibits a morphotropic phase boundary between $x=0.60$ and $0.65$ that separates a ferroelectric monoclinic \hmn{Cc} phase from a monoclinic \hmn{P2_1} phase with mixed antipolar and polar off-centering, with both the tilt pattern and the off-centering order changing discontinuously.
The composition--temperature phase diagram reproduces the known transition sequences and temperatures of both end members and reveals an intermediate tetragonal \hmn{I4/mcm} phase near the boundary that has no counterpart in either parent compound.
Above the transition temperatures the $B$ cations remain locally off-centered but orientationally disordered.
The structural transition at the boundary causes a jump in the band gap at low temperature, whereas at \qty{350}{\kelvin} the gap decreases smoothly with Sn content.
These results establish $B$-site mixing as a means to tune polar and rotational order in lead-free halide perovskites.
\end{abstract}. 

\maketitle

\section{Introduction}

Metal halide perovskites have become leading photovoltaic absorbers because they combine strong optical absorption, long carrier diffusion lengths, and defect-tolerant electronic structures \cite{stranks2013diffusion, nrel-efficiencies, Giustino2016}. The best-performing compositions, however, contain Pb, whose toxicity raises concerns for large-scale deployment \cite{babayigit2016toxicity, ke2019prospects}. The most direct replacements are Sn and Ge, whose divalent cations are isovalent with \ce{Pb^{2+}} and retain the $ns^2$ valence configuration that governs the electronic structure and lattice dynamics of the Pb compounds \cite{xiao2019lead, fu2021stereochemical, fabini2020underappreciated,ahmed2025lead}.
Removing Pb does not remove the stability problem, since both \ce{Sn^{2+}} and \ce{Ge^{2+}} are prone to oxidation \cite{noel2014lead, leijtens2017tin, krishnamoorthy2015lead}. Mixing the two cations on the $B$ site has, however, proven beneficial.
In \ce{CsSn_{0.5}Ge_{0.5}I3}, a Ge-rich native oxide passivates the perovskite surface and yields lead-free solar cells with markedly improved operational stability \cite{chen2019native}. In \ce{CsGe_{1-$x$}Sn_{$x$}Br3} films, a higher Ge content suppresses the oxidation of both cations, and the band gap can be tuned continuously between approximately \qty{1.8}{\electronvolt} and \qty{2.5}{\electronvolt} \cite{kama2022sngebr3}.

Beyond stability, Ge and Sn favor different structural distortions. \ce{CsGeBr3} is dominated by polar $B$-site off-centering. At room temperature it adopts a rhombohedral \hmn{R3m} structure, in which Ge is displaced along the pseudocubic $\langle 111 \rangle$ direction without octahedral tilting, and it becomes cubic near \qty{511}{\kelvin} \cite{thiele1987kristallstrukturen, zhang2022ferroelectricity}. The \hmn{R3m} phase is ferroelectric with an experimentally switchable polarization \cite{zhang2022ferroelectricity}, and its polar order competes closely with antipolar arrangements of the same local displacements \cite{kashikar2024coexistence}. At \qty{0}{\kelvin}, however, first-principles calculations find that octahedral tilts develop on top of the polar distortion, lowering the symmetry to a monoclinic \hmn{Cc} ground state \cite{baskurt2026polaron}. \ce{CsSnBr3} is instead dominated by octahedral tilting. On cooling, it transforms from cubic \hmn{Pm-3m} to tetragonal \hmn{P4/mbm} and orthorhombic \hmn{Pnma} near \qty{290}{\kelvin} and \qty{270}{\kelvin}, respectively \cite{mori1986x, fabini2016dynamic, fabini2024noncollinear}. Sn off-centering is present locally and dynamically in the cubic phase \cite{fabini2016dynamic, yang2017spontaneous}. It orders only at low temperature, where its coupling to the tilts produces a ferroaxial \hmn{P2_1/m} phase below approximately \qty{85}{\kelvin} and a polar, chiral \hmn{P2_1} ground state below approximately \qty{77}{\kelvin} \cite{fabini2024noncollinear}. Both end members thus combine tilting with off-centering at low temperature, but they differ in which distortion dominates.

The different balance between off-centering and tilting in the two end members has both a chemical and a geometric origin, and the two reinforce each other. The stereochemical activity of the $ns^2$ lone pair increases on ascending group 14, because the higher-lying $ns$ states couple more strongly to the halide $p$ states \cite{walsh2011stereochemistry, fabini2020underappreciated}. The Ge $4s^2$ lone pair consequently drives off-centering far more effectively than its Sn $5s^2$ counterpart \cite{stoumpos2015hybrid, radha2018distortion}. At the same time, the smaller \ce{Ge^{2+}} cation raises the Goldschmidt tolerance factor from below unity in \ce{CsSnBr3} to approximately unity in \ce{CsGeBr3}, weakening the geometric driving force for tilting \cite{goldschmidt1926gesetze, travis2016application}. Both factors therefore favor off-centering on the Ge-rich side and tilting on the Sn-rich side \cite{radha2018distortion, gao2021metal}.

Placing these competing instabilities on a common $B$-site sublattice may produce behavior that is not a simple interpolation between the end members. In oxide ferroelectrics, a \gls{mpb} occurs where two distortion patterns exchange stability, and the resulting structural degeneracy promotes intermediate phases and enhanced functional responses \cite{noheda1999monoclinic, fu2000polarization, ahart2008origin}. An analogous competition between polar and rotational order can be expected in \ce{CsGe_{1-$x$}Sn_{$x$}Br3} at intermediate compositions, with consequences for transition temperatures and local structure.

Experiments so far provide only a room-temperature view. 
X-ray diffraction finds the \hmn{R3m} structure for Ge-rich and the cubic structure for Sn-rich compositions, with a crossover between $x \approx 0.4$ and $0.6$ \cite{hooper2026synthetic}, whereas spray-deposited films retain the cubic structure of \ce{CsSnBr3} up to a Ge content of \qty{77}{\percent} \cite{kama2022sngebr3}, and the structures obtained depend on the synthesis route \cite{hooper2026synthetic}.
On the computational side, finite temperature simulations of perovskite alloys have addressed $A$ site \cite{hainer2025mpb} and halide \cite{liang2025mixedhalide} mixing. In a recent study of \ce{MAGe_{1-$x$}Sn_{$x$}I3}, we showed that Ge-driven off-centering persists up to approximately \qty{65}{\percent} Sn and constrains the orientations of the molecular $A$-site cations \cite{fransson2026magesn}. There, the structural competition is coupled to molecular reorientation. The all-inorganic \ce{CsGe_{1-$x$}Sn_{$x$}Br3} isolates the interplay between off-centering and tilting, yet no composition--temperature phase diagram has been reported for it.

Here, we determine this phase diagram using large-scale \gls{md} simulations driven by a charge-aware \gls{mlip}. We train the potential on hybrid-functional data and first test whether the predicted off-centering depends on the fraction of exact exchange.
We then use the potential-energy surfaces of the end members to examine how tilting and off-centering couple, and follow the ground-state structure as Ge is progressively replaced by Sn.
Heating simulations across the full composition range yield the composition--temperature phase diagram, which we compare with the available structural data and use to distinguish the average structure from the local displacements at the $B$ site.
Finally, we examine how these structural changes are reflected in the band gap.

\section{Methods}
\subsection{Density functional theory calculations}\label{sec:dft}

Reference data for developing the \gls{mlip} were generated using \gls{dft} calculations performed with the all-electron, numeric atom-centered orbital code \textsc{fhi-aims} \cite{blum2009ab}.
Exchange and correlation were described using the r$^2$SCAN50 hybrid meta-generalized gradient approximation \cite{Furness2020, Bursch2022}, which combines the r$^2$SCAN functional with \qty{50}{\percent} exact Hartree--Fock exchange.
This choice was motivated by studies of \ce{CsGeBr3}, in which semilocal functionals underestimated the \hmn{R3m}-to-\hmn{Pm-3m} transition temperature, whereas r$^{2}$SCAN50 reproduced the experimental value \cite{kashikar2026dft,fransson2026gecorr,Priyanka-paper}.

Scalar-relativistic effects were treated within the atomic zero-order regular approximation (atomic ZORA), and the \emph{intermediate} numeric atom-centered basis set was employed for all atomic species \cite{blum2009ab}.
Brillouin-zone integrations were performed using Monkhorst--Pack \textit{k}-point meshes automatically generated with a reciprocal-space grid density parameter of 5.6, corresponding to a \textit{k}-point spacing of approximately \qty{0.18}{\per\angstrom}.
The electronic self-consistent field calculations were converged to a charge-density threshold of \qty{e-6}{\elementarycharge\per\bohr\cubed}.
All \gls{dft} calculations consisted of single-point evaluations of pre-generated atomic configurations, whose construction is described in \autoref{sec:qnep}.
Total energies and atomic forces were computed for all 348 configurations in the final reference dataset.
Due to the substantially higher memory requirements of the hybrid r$^2$SCAN50 calculations, virial stress tensors were evaluated only for the 112 configurations containing 40 atoms or fewer.
In addition, \gls{bec} tensors were calculated for 11 representative structures, selected to span the compositional and structural diversity of the dataset.
These tensors serve as training targets for the charge model of the \gls{qnep} (\autoref{sec:qnep}), which represents long-range electrostatics through environment-dependent atomic charges.

Separately, the sensitivity of the predicted structural distortions to the exact-exchange fraction is assessed explicitly in \autoref{sec:A}.
The structures of \ce{Cs$B$Br3} ($B$~=~Ge, Sn, Pb) were relaxed in 40-atom cells using r$^2$SCAN with exact-exchange fractions of $\alpha = \qtylist{0;10;25;30;40;50}{\percent}$, where $\alpha = \qtylist{0;25;50}{\percent}$ correspond to r$^2$SCAN, r$^2$SCAN0, and r$^2$SCAN50, respectively.
The starting structures were taken from the candidate structures generated from the cubic perovskite by group-theoretical methods (\autoref{sec:C}) \cite{stokes2002grouptheory}: $R3m$ for \ce{CsGeBr3}, with Ge displaced along $\langle 111 \rangle$, and $Pnma$ ($a^-a^-c^+$) for \ce{CsSnBr3} and \ce{CsPbBr3}. Atomic positions and lattice vectors were optimized without symmetry constraints until the maximum residual force component was below \qty{e-4}{\electronvolt\per\angstrom}.
For \ce{CsSnBr3} and \ce{CsPbBr3}, the relaxations were started from the tilted $Pnma$ structure both without and with $B$-site off-centering, to ensure that the final structures do not depend on the starting point.

\subsection{Machine-learned interatomic potential}\label{sec:qnep}

A \gls{qnep} \cite{fan2026qnep} was constructed using the \gpumd{} package (version 4.8) \cite{fan2021nep, fan2022gpumd} with the default hyperparameters.
The training dataset consisted of 348 \gls{dft} reference configurations spanning the configurational space of \ce{CsGe_{1-$x$}Sn_{$x$}Br3}, including diverse Sn/Ge cation arrangements, crystallographic phases, and thermally distorted structures sampled from \gls{md} simulations. The configurations contain between 5 and 320 atoms, corresponding to supercells of 1 to 64 formula units, amounting to a total of 32,935 atoms, and cover Sn concentrations across the full range $0 \leq x \leq 1$.

The \gls{qnep} model was optimized by simultaneously fitting the \gls{dft} total energies, atomic forces, virial stress tensors, and \gls{bec} tensors described in \autoref{sec:dft}.
The convergence of the optimization was monitored through the evolution of the total loss together with the corresponding \glspl{rmse} for the energies, forces, virials, and \glspl{bec} (\autoref{sfig:loss_curves}).
The final \gls{qnep} reproduces the \gls{dft} training data with \glspl{rmse} of \qty{1.7}{\milli\electronvolt\per\atom} for the energies, \qty{60.3}{\milli\electronvolt\per\angstrom} for the forces, \qty{13.0}{\milli\electronvolt\per\atom} for the virials, and \qty{48.4}{\mega\pascal} for the stresses (\autoref{stab:qnep_errors}).
The associated parity plots are provided in \autoref{sfig:parity_all}, while species-resolved force parity plots are shown in \autoref{sfig:parity_species}.
The predicted \glspl{bec} also agree well with the \gls{dft} reference values, with an \gls{rmse} of \qty{0.071}{\elementarycharge} (\autoref{sfig:parity_all}).
The model therefore captures the environment-dependent charge response that underlies the long-range electrostatic interactions in halide perovskites.

To test how well the model predicts structures that were not used in training, we performed five-fold cross-validation, in which each model was trained on 313 structures and tested on the remaining 35.
The loss curves converge similarly for all five splits (\autoref{sfig:loss_curves_splits}), and the training errors are nearly identical to those of the final model (\autoref{stab:qnep_errors}).
Averaged over the five splits, the test \glspl{rmse} are \qty{3.1}{\milli\electronvolt\per\atom} for the energies, \qty{72.3}{\milli\electronvolt\per\angstrom} for the forces, and \qty{0.074}{\elementarycharge} for the \glspl{bec}, only moderately above the training errors.
The larger mean test errors for the virials and stresses arise from a single high-energy structure in the test set of one split (\autoref{stab:qnep_errors_splits}).
The model thus generalizes well to unseen structures within the configurational space covered by the training data.

\subsection{Molecular dynamics simulations}\label{sec:md}

Finite-temperature properties of \ce{CsGe_{1-$x$}Sn_{$x$}Br3} were investigated using large-scale \gls{md} simulations performed with the \gls{qnep} model implemented in the \gpumd{} package.
All simulations were carried out in the isothermal--isobaric ($NPT$) ensemble using the stochastic cell-rescaling thermostat and barostat \cite{bernetti2020pressure} as implemented in \gpumd{}, allowing both the atomic positions and simulation cell to evolve during the simulations.

Supercells containing \num{40000} atoms, obtained by repeating the 5000-atom ground-state cells of \autoref{sec:C} $2\times2\times2$ times, were employed for all production simulations to minimize finite-size effects.
The equations of motion were integrated using a time step of \qty{5}{\femto\second}.

To determine the phase diagram, heating simulations were started from the ground-state structure of each composition (\autoref{sec:C}).
The temperature was increased linearly from \qty{1}{\kelvin} to \qty{600}{\kelvin} over \num{4e7} steps, corresponding to \qty{200}{\nano\second}.
Lattice parameters and angles, octahedral tilt angles, and $B$-site off-centering displacements were evaluated along the trajectories to identify the structural phase transitions.
Octahedral tilt angles were computed from the positions of the Br atoms of each octahedron following Ref.~\citenum{wiktor2023tilting}, and the off-centering displacement of each $B$ cation was measured relative to the center of its surrounding octahedron as in Ref.~\citenum{fransson2026magesn}.
Transition temperatures obtained from heating runs can be affected by superheating, so the values reported here should be regarded as estimates and upper bounds \cite{fransson2023phase}.

\subsection{Band gap calculations}\label{sec:bandgap}

Band gaps were computed with \gls{dft} using the PBEsol functional \cite{perdew2008pbesol} in \vasp{} \cite{kresse1996efficient}, chosen for its low cost, which allows many configurations to be sampled.
The structures were 80-atom supercells of the Ge and Sn ground-state prototypes with random Sn/Ge occupations, up to 30 per composition, relaxed with the \gls{qnep}.
The same structures were used to start \gls{md} simulations at \qty{350}{\kelvin} with the \gls{qnep} in 160-atom cells, from which snapshots were taken.
Single-point calculations on the relaxed structures and on the snapshots then gave the band gaps, without spin--orbit coupling.
These calculations are meant to capture the qualitative changes of the band gap with composition and temperature, not accurate absolute values.

\section{Results and discussion}

\subsection{Order parameters and the effect of exact exchange}\label{sec:A}

\begin{figure}[t]
    \includegraphics[scale=0.055]{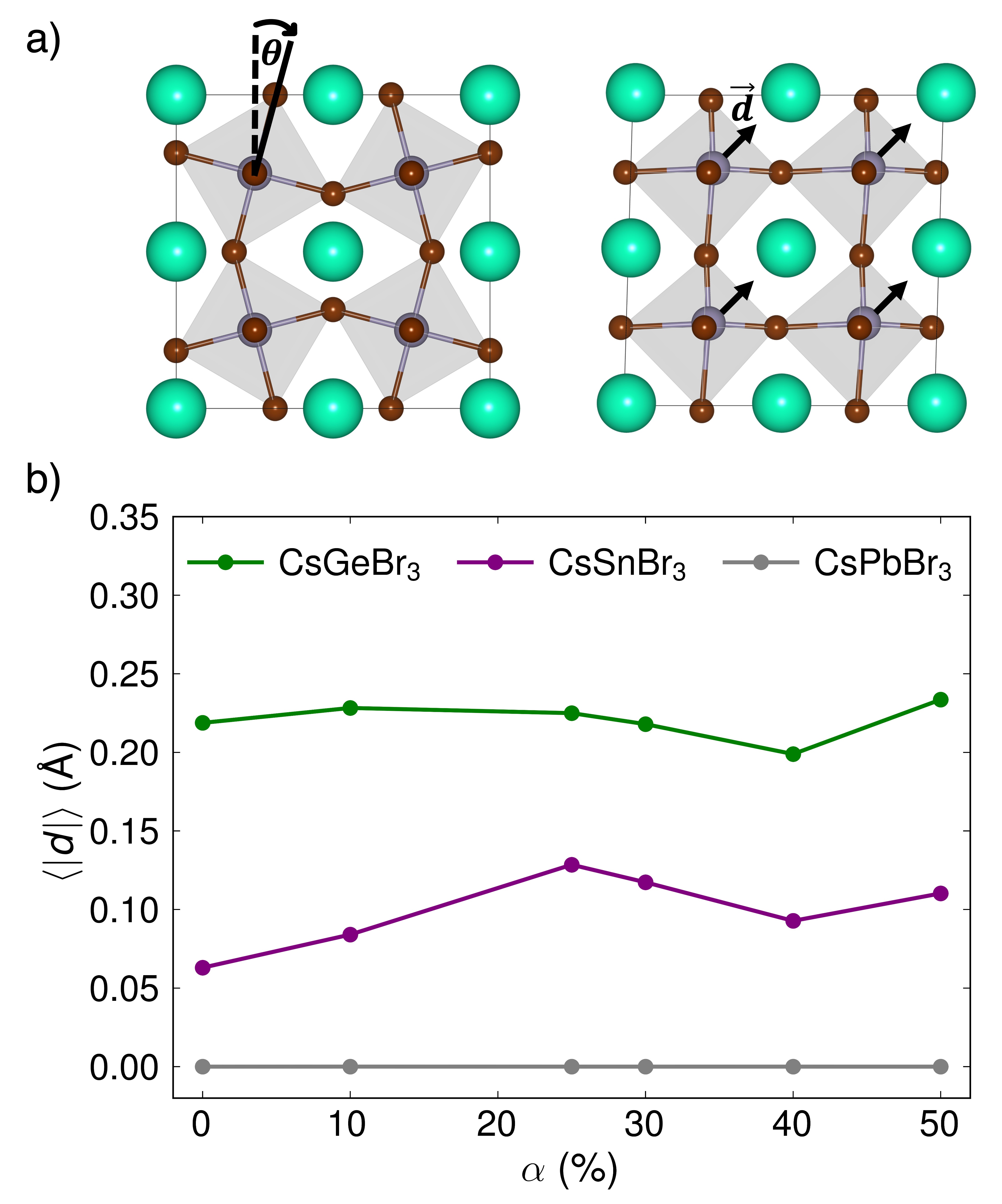}
    \caption{(a) Schematic of the two order parameters used in this work, illustrated for \ce{Cs$B$Br3} ($B$~=~Ge, Sn, Pb): the octahedral tilt angle $\theta$ (left) and the $B$-site off-centering displacement $\vec{d}$ (right), indicated by the black arrows.
    Green, light purple, and brown spheres denote Cs, $B$-site, and Br atoms, respectively.
    (b) Average $B$-site off-centering $\langle |d| \rangle$ as a function of the exact-exchange fraction $\alpha$ in the r$^2$SCAN$\alpha$ hybrid functional for \ce{CsGeBr3}, \ce{CsSnBr3}, and \ce{CsPbBr3}, relaxed from \hmn{R3m} (Ge) and tilted \hmn{Pnma} (Sn, Pb) starting structures without symmetry constraints (\autoref{sec:dft}). Here, $\langle |d| \rangle$ is the average of the absolute values of the Cartesian components of the off-centering displacement over all $B$ atoms, so that antipolar arrangements contribute in the same way as polar ones.}
    \label{fig:structures-disp-alpha}
\end{figure}

The structural distortions in \ce{Cs$B$Br3} ($B$~=~Ge, Sn, Pb) are described throughout this work by two order parameters, illustrated in \autoref{fig:structures-disp-alpha}a.
The octahedral tilt angle $\theta$ measures the rotation of the $B$Br$_6$ octahedra relative to the ideal cubic structure, and the $B$-site off-centering displacement $d$ measures the displacement of the $B$ cation from the center of its surrounding octahedron.
These are the two principal distortion types whose superposition determines the symmetry of a distorted perovskite \cite{howard1998tilting, stokes2002grouptheory}.
Octahedral tilting transforms as the $M_3^+$ (in-phase) and $R_4^+$ (out-of-phase) irreducible representations of \hmn{Pm-3m}, and ferroelectric $B$-cation displacement as $\Gamma_4^-$, whereas antipolar displacement patterns, in which neighboring $B$ cations move in opposite directions, correspond to zone-boundary representations.

Structures are labeled throughout using the modified Glazer notation of Stokes \emph{et al.} \cite{glazer1972tilt, stokes2002grouptheory}, in which the three letters and their superscripts denote the tilt system in the usual way and a subscript records the ordering of the $B$-cation displacement along each axis.
We adapt this scheme to the phases encountered here.
Ferroelectric ordering of the displacements along an axis is denoted by the subscript FE, antiferroelectric ordering by AFE, and dynamically disordered off-centering, in which the local displacements remain finite but their average vanishes, by DDO (\autoref{sec:D}).
Thus $a^{-}_{\mathrm{FE}}a^{-}_{\mathrm{FE}}c^{-}_{\mathrm{FE}}$ denotes out-of-phase tilts about all three axes with ferroelectrically ordered displacements along each, and $a^{0}_{\mathrm{DDO}}a^{0}_{\mathrm{DDO}}a^{0}_{\mathrm{DDO}}$ an untilted structure whose $B$ cations are locally displaced but orientationally disordered.

The reference data for the \gls{mlip} were computed with the r$^2$SCAN50 hybrid functional (\autoref{sec:dft}) because semilocal functionals do not describe the structural energetics of these materials reliably.
Transition temperatures of halide perovskites depend strongly on the exchange--correlation functional \cite{fransson2023phase}, and for \ce{CsGeBr3} r$^2$SCAN50 reproduces the experimental \hmn{R3m}-to-\hmn{Pm-3m} transition temperature, which semilocal functionals underestimate \cite{kashikar2026dft}. Exact exchange, however, also changes the relative energies of the $B$-$s$ and Br-$p$ states whose interaction drives the lone-pair distortion \cite{walsh2011stereochemistry}, and it could therefore exaggerate the off-centering or create it where none exists.
Before turning to the mixed system, we test this by relaxing the two end members, \ce{CsGeBr3} and \ce{CsSnBr3}, together with \ce{CsPbBr3}, with exact-exchange fractions between $\alpha = \qty{0}{\percent}$ (r$^2$SCAN) and \qty{50}{\percent} (r$^2$SCAN50), starting from \hmn{R3m} for \ce{CsGeBr3} and from \hmn{Pnma} for \ce{CsSnBr3} and \ce{CsPbBr3} (\autoref{sec:dft}).
\ce{CsPbBr3} serves as a reference case: the Pb $6s^2$ lone pair is the least stereochemically active of the group-14 cations, and the Pb site is known to remain centrosymmetric \cite{frohna2018inversion, fabini2020underappreciated}, so any off-centering obtained for Pb would reveal an artifact of exact exchange.

\autoref{fig:structures-disp-alpha}b shows the resulting average off-centering $\langle |d| \rangle$.
In \ce{CsGeBr3}, the structure remains $R3m$ and $\langle |d| \rangle$ stays between \qty{0.20}{\angstrom} and \qty{0.23}{\angstrom} for all $\alpha$, with no systematic trend.
In \ce{CsSnBr3}, the Sn cations move off the inversion centers of the $Pnma$ starting structure at every $\alpha$, so that the relaxed structure has lower symmetry than $Pnma$, consistent with the off-centered low-temperature phases observed experimentally \cite{fabini2024noncollinear}.
The displacement is smaller than in \ce{CsGeBr3} and depends more on $\alpha$, ranging from \qty{0.06}{\angstrom} with r$^2$SCAN to \qtyrange{0.09}{0.13}{\angstrom} with hybrid functionals ($\alpha \geq \qty{25}{\percent}$).
By contrast, \ce{CsPbBr3} remains essentially centrosymmetric for all values of $\alpha$, exhibiting negligible off-centering throughout. For both \ce{CsSnBr3} and \ce{CsPbBr3}, relaxations started with and without $B$-site off-centering converged to the same structures.

Exact exchange therefore does not create off-centering where none is expected. Pb remains centrosymmetric,
whreas Ge and Sn are displaced already with semilocal r$^2$SCAN, with magnitudes following the expected lone-pair trend, Ge $>$ Sn $>$ Pb, at every $\alpha$ \cite{walsh2011stereochemistry, fu2021stereochemical}.
This is consistent with experiment, which finds off-centered Ge in $R3m$ \ce{CsGeBr3} \cite{thiele1987kristallstrukturen, zhang2022ferroelectricity} and ordered Sn off-centering in \ce{CsSnBr3} below approximately \qty{85}{\kelvin} \cite{fabini2024noncollinear}.
In \ce{CsGeBr3}, exact exchange acts mainly on the energetics. The energy of the relaxed \hmn{R3m} structure relative to cubic \hmn{Pm-3m} decreases from \qty{-30.0}{meV/atom} with r$^2$SCAN0 to \qty{-39.1}{meV/atom} with r$^2$SCAN50, although the Ge displacement hardly changes.
Exact exchange thus deepens the off-centering energy well rather than enlarging the distortion, consistent with the improved transition temperature of \ce{CsGeBr3} obtained with r$^2$SCAN50 \cite{kashikar2026dft,Priyanka-paper}.
In \ce{CsSnBr3}, exact exchange affects both, enlarging the displacement relative to r$^2$SCAN and possibly also overstabilizing the off-centering.
Since a single functional must describe the whole composition range, r$^2$SCAN50 is thus a reasonable compromise. It is chosen to describe the Ge-driven off-centering accurately, while still reproducing the tilt transitions of \ce{CsSnBr3} well (\autoref{sec:D}), although a possible overestimation of Sn off-centering may contribute to the higher predicted ordering temperature.

\subsection{Energetic interplay between octahedral tilting and \texorpdfstring{$B$}{B}-site off-centering}\label{sec:B}

To gain microscopic insight into the interplay between the two distortions, we mapped the \glspl{pes} associated with the relevant lattice modes using the \gls{qnep}.
The $B$-site off-centering is described by the ferroelectric mode amplitude $Q_\mathrm{FE\langle111\rangle}$, corresponding to polar displacement along the pseudocubic $\langle 111 \rangle$ direction; the octahedral tilting by the in-phase $M$-point mode $Q_M$ and the out-of-phase $R$-point mode $Q_R$.
The mode eigenvectors were obtained from harmonic phonon calculations for the cubic phase in $2\times2\times2$ supercells using \phonopy{} \cite{togo2023phonopy} and projected out using the mode projection functionality of \dynasor{} \cite{fransson2021dynasor, berger2025dynasor}.
One- and two-dimensional surfaces were constructed by varying the amplitudes of these modes while keeping all other degrees of freedom fixed, thereby isolating the energetic coupling between the distortions.

\begin{figure*}[t]
    \includegraphics{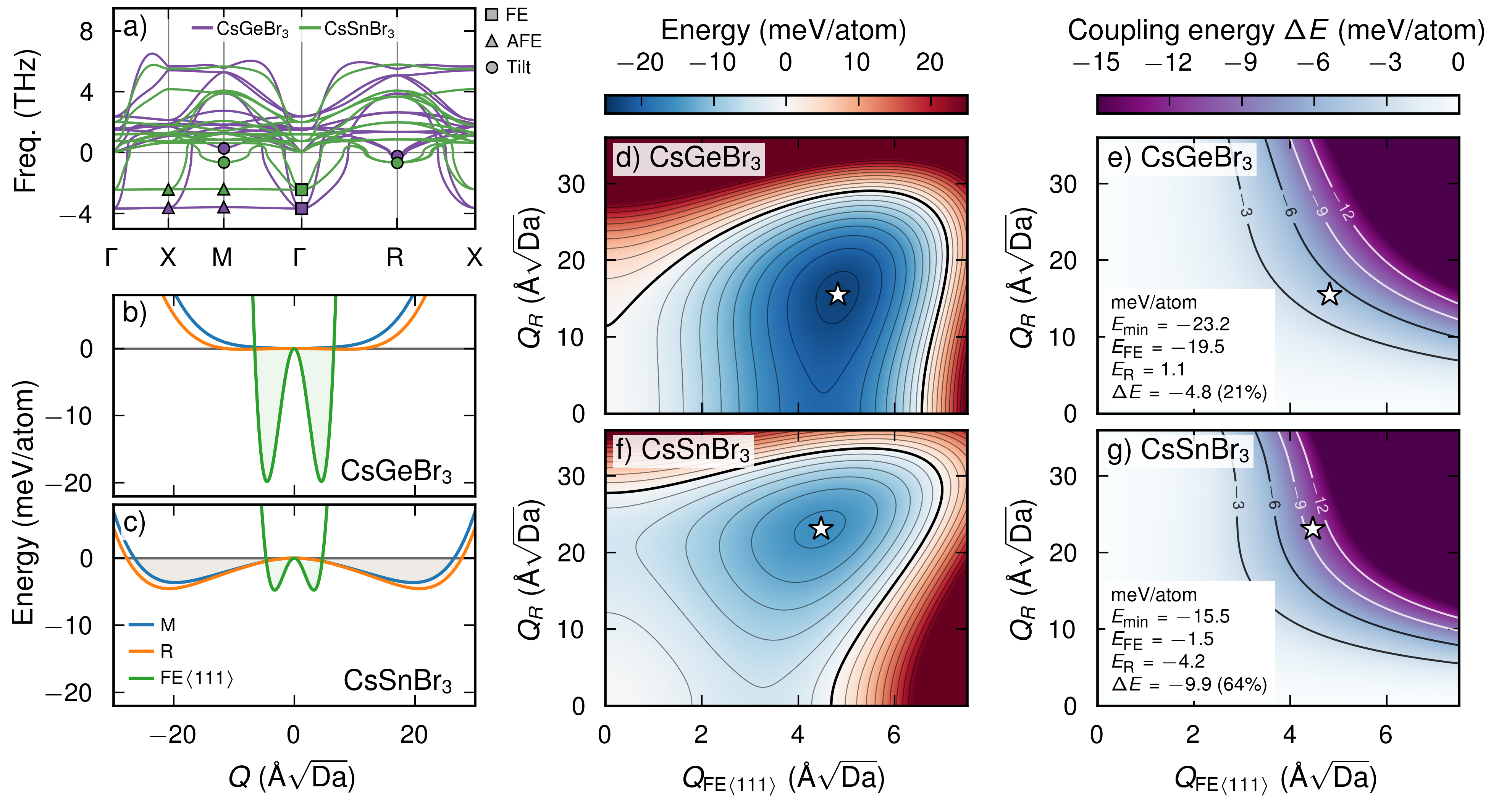}
\caption{Lattice instabilities and mode coupling in cubic \ce{CsGeBr3} (b, d, e) and \ce{CsSnBr3} (c, f, g).
(a) Harmonic phonon dispersions.
Markers indicate the ferroelectric mode at $\Gamma$ (squares), the antiferroelectric modes at X and M (triangles), and the tilt modes at M and R (circles).
(b,c) Energy as a function of the amplitude of the ferroelectric mode $Q_{\mathrm{FE}\langle 111\rangle}$, the in-phase tilt mode $Q_M$, and the out-of-phase tilt mode $Q_R$.
(d,f) Energy as a function of $Q_{\mathrm{FE}\langle 111\rangle}$ and $Q_R$, with stars marking the minimum.
(e,g) Coupling energy $\Delta E$ according to \autoref{eq:coupling}.
The boxes give the energy at the minimum, $E_\mathrm{min} = E_\mathrm{FE} + E_\mathrm{R} + \Delta E$, with the share of $\Delta E$ in parentheses.
Contour lines are spaced by \qty{3}{\milli\electronvolt\per\atom} and the thick line marks zero.
Energies are given relative to the cubic structure.}
    \label{fig:mode}
\end{figure*}

\autoref{fig:mode}a shows the harmonic phonon dispersions of the cubic phases.
Both compounds have unstable modes of three kinds: octahedral tilting at the R and M points, antiferroelectric $B$-site displacements at the X and M points, and the ferroelectric $B$-site displacement at $\Gamma$.
The one-dimensional \glspl{pes} in \autoref{fig:mode}b,c show how strongly these instabilities differ between the two compounds.
In \ce{CsGeBr3} the ferroelectric mode forms a deep double well of about \qty{20}{\milli\electronvolt\per\atom}, whereas the $M$ tilt mode is stable and the $R$ tilt mode is only marginally unstable.
In \ce{CsSnBr3} the ferroelectric well is much shallower and comparable in depth to the tilt wells, which are now clearly unstable.
This difference reflects the weaker stereochemical activity of the Sn $5s^2$ lone pair compared to the Ge $4s^2$ lone pair \cite{walsh2011stereochemistry, radha2018distortion}, consistent with the off-centering in \autoref{fig:structures-disp-alpha}b.

To examine how the two kinds of distortion interact, \autoref{fig:mode}d,f show the \gls{pes} as a function of $Q_\mathrm{FE\langle111\rangle}$ and $Q_R$, taking the $\langle 111 \rangle$ off-centering and the out-of-phase tilt as representative modes.
It has been argued that tilting and off-centering are decoupled by symmetry \cite{hyltonfarrington2026notcoupled}.
A bilinear coupling is indeed symmetry forbidden, but higher-order coupling is allowed, and the \glspl{pes} show that it matters.
In \ce{CsGeBr3} the $R$ tilt alone is barely unstable, yet once the ferroelectric distortion is present, tilting lowers the energy further and the joint minimum (star) lies at a sizable tilt amplitude.
In \ce{CsSnBr3} the two modes also reinforce each other, and the joint minimum lies well below the minima of the two one-dimensional \glspl{pes}.
Note that these minima involve only the two modes considered here, a typical tilting and ferroelectric mode, whereas the actual ground states (\autoref{sec:C}) contain additional distortions.

To quantify the interaction, \autoref{fig:mode}e,g show the coupling energy
\begin{align}
    \Delta E = E(Q_\mathrm{FE\langle111\rangle}, Q_R) - E_\mathrm{FE} - E_\mathrm{R},
    \label{eq:coupling}
\end{align}
where $E_\mathrm{FE} = E(Q_\mathrm{FE\langle111\rangle})$ and $E_\mathrm{R} = E(Q_R)$ are the one-dimensional \glspl{pes}, i.e., $\Delta E$ is the part of the energy that is not captured by them.
In both compounds $\Delta E$ is negative, but its magnitude differs strongly.
At the joint minimum the coupling contributes \qty{4.8}{\milli\electronvolt\per\atom} (\qty{21}{\percent} of the total energy gain) in \ce{CsGeBr3} but \qty{9.9}{\milli\electronvolt\per\atom} (\qty{64}{\percent}) in \ce{CsSnBr3}.
In \ce{CsGeBr3} the energy gain is thus dominated by the ferroelectric mode itself, whereas in \ce{CsSnBr3} it arises mainly from the coupling between off-centering and tilting.
In the language of a Landau expansion, the mode-coupling terms are therefore much stronger in \ce{CsSnBr3} than in \ce{CsGeBr3}.
This sets up the composition-dependent competition examined next: replacing Ge by Sn weakens the polar instability while strengthening both the tilt instabilities and their coupling to the off-centering.

\subsection{Composition-dependent ground-state structures}\label{sec:C}

\begin{figure*}[t]
    \includegraphics[scale=0.96]{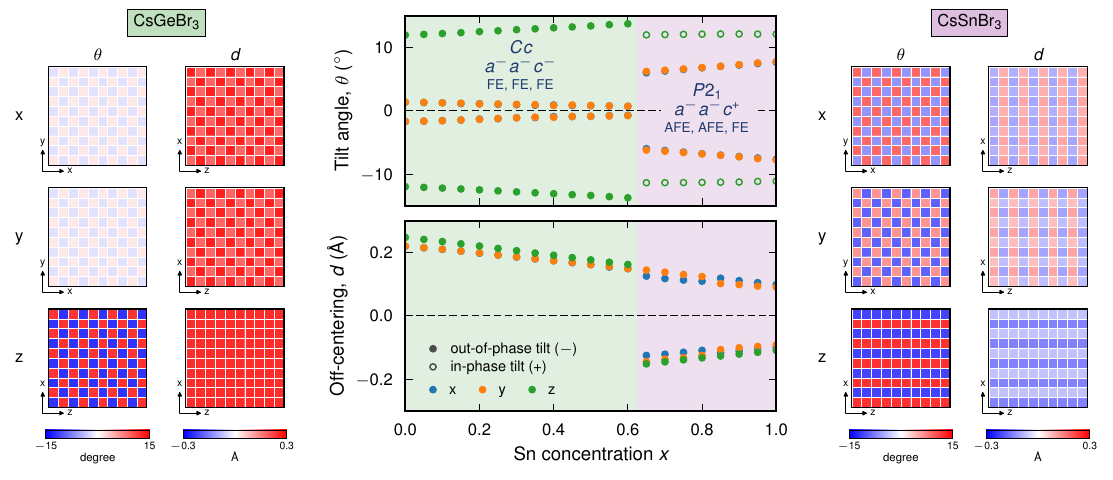}
    \caption{Ground-state octahedral tilt angle ($\theta$) and $B$-site off-centering displacement ($d$) in \ce{CsGe_{1-$x$}Sn_{$x$}Br3}.
    The left and right panels show representative maps of $\theta$ and $d$ along the $x$, $y$, and $z$ directions for \ce{CsGeBr3} and \ce{CsSnBr3}, respectively.
    The center panels show the composition dependence of the averaged positive and negative branches of the $\theta$ and $d$ components along the $x$, $y$, and $z$ directions.
    A ground-state phase boundary is identified between $x = 0.60$ and $0.65$, separating the monoclinic \hmn{Cc} phase with an $a^-a^-c^-$ tilt pattern and FE, FE, FE off-centering in the Ge-rich (green) region from the monoclinic \hmn{P2_1} phase with an $a^-a^-c^+$ tilt pattern and AFE, AFE, FE off-centering in the Sn-rich (purple) region.
    Here, FE and AFE denote ferroelectric and antiferroelectric ordering of the off-centering, respectively.}
    \label{fig:ground_states}
\end{figure*}

The ground-state structures of \ce{CsGe_{1-$x$}Sn_{$x$}Br3} were determined using the \gls{qnep} by relaxing 61 candidate perovskite polymorphs generated with group-theoretical methods \cite{stokes2002grouptheory}, spanning a wide range of octahedral tilt patterns and polar distortions.
For each composition a randomized Sn/Ge distribution was generated in a 5000-atom supercell, which is large enough to host both chemical disorder and the tilt and off-centering patterns.
The same randomized configuration was used as the starting point for all 61 candidates, both atomic positions and cell were relaxed, and the relaxed structure of lowest total energy was taken as the ground state.
Since the $B$-site occupation is random, the space group was assigned from the tilt and off-centering patterns rather than from the atomic positions.
Notably, the antipolar arrangement of the Sn-rich ground state emerged during relaxation although none of the 61 candidates contains such a pattern.
The resulting structures were characterized using the order parameters $\theta$ and $d$ of \autoref{sec:A}.
This treatment of the $B$-site sublattice as a random solid solution is consistent with the experimental evidence for homogeneous alloy formation \cite{kama2022sngebr3}.

The calculated ground-state structures reveal a sharp composition-driven structural transition between $x = 0.60$ and $0.65$ (\autoref{fig:ground_states}).
Ge-rich compositions ($x \leq 0.60$) adopt a monoclinic \hmn{Cc} structure with an $a^-a^-c^-$ tilt pattern and ferroelectric ordering of the $B$-site off-centering along all three crystallographic directions (FE, FE, FE).
Sn-rich compositions ($x \geq 0.65$) instead stabilize a monoclinic \hmn{P2_1} phase with an $a^-a^-c^+$ tilt pattern, in which the off-centering becomes antiferroelectric along $x$ and $y$ while remaining ferroelectric along $z$ (AFE, AFE, FE). The \hmn{Cc} ground state of \ce{CsGeBr3} agrees with recent hybrid-functional calculations, which also find a monoclinic \hmn{Cc} structure combining Ge off-centering with out-of-phase tilts \cite{baskurt2026polaron}, and the \hmn{P2_1} symmetry of the Sn-rich ground state matches that assigned experimentally to \ce{CsSnBr3} below \qty{77}{\kelvin} \cite{fabini2024noncollinear}.

The evolution of the order parameters across this boundary illustrates the competition directly.
Within the \hmn{Cc} phase the off-centering magnitude decreases gradually with increasing Sn content, tracking the progressive weakening of the lone-pair distortion as Ge is replaced.
The tilt pattern is dominated by a large out-of-phase tilt about $z$, which grows slightly with Sn content, whereas the out-of-phase tilts about $x$ and $y$ remain small. The distortion pattern itself remains robust and only its amplitude responds to composition.

At the boundary both order parameters change discontinuously.
The large out-of-phase tilt about $z$ changes to an in-phase tilt of similar magnitude, while the two small out-of-phase tilts about $x$ and $y$ increase to about \ang{6} and continue to grow with Sn content, producing the $a^-a^-c^+$ pattern of the \hmn{P2_1} phase.
At the same time, the off-centering decreases slightly and reorganizes into an antiferroelectric arrangement along $x$ and $y$, while remaining ferroelectric along $z$.
The \hmn{Cc} and \hmn{P2_1} phases are not related by a group--subgroup relation, nor are their tilt systems $a^-a^-c^-$ and $a^-a^-c^+$ \cite{howard1998tilting, stokes2002grouptheory}, so the transition between the two monoclinic ground states cannot be continuous and is first order.

\subsection{Finite-temperature phase transitions}\label{sec:D}
\begin{figure*}[t]
    \includegraphics{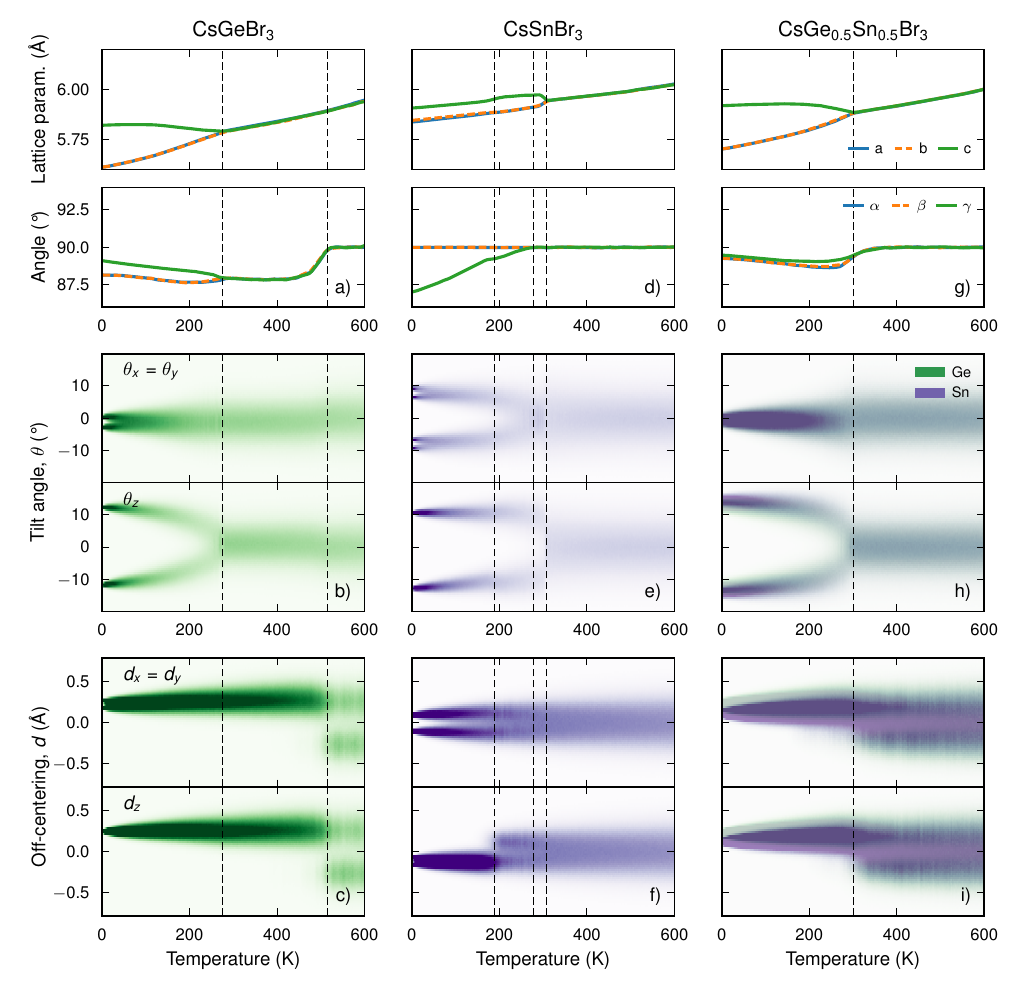}
  \caption{Finite-temperature structural evolution of \ce{CsGeBr3} (a--c), \ce{CsSnBr3} (d--f), and the equiatomic mixed composition \ce{CsGe_{0.5}Sn_{0.5}Br3} (g--i), obtained from \gls{md} simulations using the \gls{qnep}. (a,d,g) Temperature dependence of the lattice parameters ($a$, $b$, and $c$) and lattice angles ($\alpha$, $\beta$, and $\gamma$). (b,e,h) Probability distributions of the octahedral tilt angles along the pseudocubic $x$, $y$ ($\theta_x = \theta_y$), and $z$ ($\theta_z$) directions as a function of temperature. (c,f,i) Corresponding probability distributions of the $B$-site off-centering displacements along the pseudocubic $x$, $y$ ($d_x = d_y$), and $z$ ($d_z$) directions.
  Throughout, green and purple denote quantities resolved on Ge- and Sn-centered octahedra, respectively, so that the two contributions appear separately for the mixed composition in h) and i).
  Vertical dashed lines indicate the structural phase transition temperatures.
}
    \label{fig:phase-transition}
\end{figure*}

We next follow the evolution of these ground states with temperature using \gls{md} simulations driven by the \gls{qnep}.
Heating runs were started from the ground-state structure of each composition identified in \autoref{sec:C} (\autoref{sec:md}). \autoref{fig:phase-transition}a--c show the behavior of \ce{CsGeBr3}.
At low temperature the system adopts the monoclinic \hmn{Cc} ground state with the $a^-_\mathrm{FE}a^-_\mathrm{FE}c^-_\mathrm{FE}$ distortion pattern, characterized by unequal lattice parameters, non-orthogonal lattice angles, finite octahedral tilts, and large ferroelectric off-centering along all three directions.
The first transition occurs near \qty{275}{\kelvin}, where the monoclinic distortion disappears and the lattice becomes rhombohedral (\hmn{R3m}).
The out-of-phase tilts are suppressed while finite off-centering persists along all three directions, giving the $a^0_\mathrm{FE}a^0_\mathrm{FE}a^0_\mathrm{FE}$ pattern.
The increase in crystal symmetry is thus driven by the loss of octahedral tilting, not by the loss of ferroelectric order.

A second transition occurs near \qty{515}{\kelvin}, where the average tilts and the average off-centering both vanish and the structure becomes cubic (\hmn{Pm-3m}).
The off-centering distributions (\autoref{fig:phase-transition}c) show, however, that the Ge atoms remain locally displaced from the centrosymmetric positions even though the average polar distortion is zero.
Rather than becoming statically centrosymmetric, the high-temperature cubic phase fluctuates dynamically between symmetry-equivalent polar configurations.
We describe this state as exhibiting dynamically disordered off-centering (DDO), corresponding to the $a^{0}_{\mathrm{DDO}}a^{0}_{\mathrm{DDO}}a^{0}_{\mathrm{DDO}}$ pattern.
The correlations between the local Ge displacements in this phase are analyzed in detail in Ref.~\citenum{fransson2026gecorr}.
This is consistent with the dynamic Sn off-centering observed experimentally in cubic \ce{CsSnBr3} \cite{fabini2016dynamic}.

The predicted sequence for \ce{CsGeBr3} is thus \hmn{Cc} $\rightarrow$ \hmn{R3m} $\rightarrow$ \hmn{Pm-3m}.
The predicted \hmn{R3m} $\rightarrow$ \hmn{Pm-3m} transition at \qty{515}{\kelvin} agrees closely with the experimental value of \qty{511}{\kelvin} \cite{thiele1987kristallstrukturen, zhang2022ferroelectricity}, whereas the low-temperature \hmn{Cc} $\rightarrow$ \hmn{R3m} transition has not been reported experimentally.
The predicted \qty{275}{\kelvin} is also higher than the approximately \qty{160}{\kelvin} obtained with an r$^2$SCAN50 potential trained on \ce{CsGeBr3} alone \cite{Priyanka-paper}.
Both potentials reproduce the \hmn{R3m} $\rightarrow$ \hmn{Pm-3m} transition but differ in the small energy gained by tilting in the polar phase.
Relaxed with the \gls{qnep}, \hmn{R3m} lies \qty{5.5}{\milli\electronvolt\per\atom} above \hmn{Cc}, whereas r$^2$SCAN50 gives about \qty{1}{\milli\electronvolt\per\atom} \cite{Priyanka-paper}.
The \gls{qnep} thus overestimates the tilt stabilization in Ge-rich environments.
Additionally, superheating (\autoref{sec:md}) and the neglect of nuclear quantum effects, which tend to suppress such shallow instabilities, could possibly lead to overestimation of the transition temperature.

\ce{CsSnBr3} behaves quite differently (\autoref{fig:phase-transition}d--f), as expected from the dominance of octahedral tilting.
The low-temperature monoclinic \hmn{P2_1} ground state has the $a^-_\mathrm{AFE}a^-_\mathrm{AFE}c^+_\mathrm{FE}$ pattern, with antiferroelectric off-centering along $x$ and $y$ and ferroelectric alignment along $z$.
The first transition occurs near \qty{189}{\kelvin} into the orthorhombic \hmn{Pnma} phase. The long-range antiferroelectric order is lost, but finite local Sn off-centering persists along all three directions, giving a dynamically disordered state ($a^-_\mathrm{DDO}a^-_\mathrm{DDO}c^+_\mathrm{DDO}$).
Experimentally, two closely spaced low-temperature transitions are observed in \ce{CsSnBr3} at \qtylist{77;85}{\kelvin}, and phonon-mode mapping led Fabini \emph{et al.} to assign the intervening phase \hmn{P2_1/m} symmetry, distinguished from \hmn{P2_1} by a noncollinear arrangement of the Sn lone-pair distortions that preserves inversion symmetry \cite{fabini2024noncollinear}.
Our simulations predict a direct \hmn{P2_1} $\rightarrow$ \hmn{Pnma} transition and do not resolve this intermediate phase.
This is not unexpected, since the two phases differ only in the relative orientation of the lone-pair displacements at a fixed tilt system and are stable experimentally over a window of less than \qty{10}{\kelvin}, which implies a very small free-energy difference between them.
Our predicted transition temperature of \qty{189}{\kelvin} is considerably higher than the experimental values.
Several factors may contribute: superheating in the heating simulations, a possible overstabilization of the Sn off-centering by exact exchange (\autoref{sec:A}), and the neglect of nuclear quantum effects, which are expected to lower transition temperatures most strongly for low-temperature transitions with shallow energy wells \cite{zhong1996quantum}.

On further heating the out-of-phase tilts disappear near \qty{277}{\kelvin}, leaving the in-phase rotation about $z$ ($a^0a^0c^+$, \hmn{P4/mbm}), and the remaining rotation vanishes near \qty{309}{\kelvin}, producing the dynamically disordered cubic phase.
The predicted transition temperatures compare favorably with the experimental values of approximately \qty{270}{\kelvin} for \hmn{Pnma} $\rightarrow$ \hmn{P4/mbm} and \qty{290}{\kelvin} for \hmn{P4/mbm} $\rightarrow$ \hmn{Pm-3m} \cite{mori1986x,fabini2016dynamic,fabini2024noncollinear}.
The sequence is therefore governed by the progressive suppression of octahedral tilting, while the local Sn lone-pair distortions remain dynamically active over the whole temperature range.

For the mixed system, representative results for the equiatomic composition, \ce{CsGe_{0.5}Sn_{0.5}Br3}, are shown in \autoref{fig:phase-transition}g--i.
Unlike the two end members, the mixed composition exhibits a single structural transition near \qty{302}{\kelvin} from the monoclinic \hmn{Cc} phase directly to the dynamically disordered cubic \hmn{Pm-3m} phase.
Both the octahedral tilts and the long-range ferroelectric ordering disappear simultaneously, while broad $B$-site off-centering distributions persist above the transition temperature, indicating that local polar distortions remain dynamically active.
The absence of the intermediate rhombohedral phase shows that, at this composition, the long-range polar order is no longer stable once the tilts are lost, which is consistent with the cooperative coupling between the two distortions found in \autoref{sec:B}. 
This conclusion, however, depends on the predicted tilt transition, which stays near \qty{300}{\kelvin} throughout the \hmn{Cc} range, whereas the loss of polar order falls steeply with increasing Sn content.
If the tilt transition is lowered more than the loss of polar order, either because the tilts are overstabilized by the \gls{qnep}, as in \ce{CsGeBr3}, or because it is more strongly affected by superheating (\autoref{sec:md}), an \hmn{R3m} range could appear at this composition.
Because the polar order weakens steadily with Sn content, such a range would nevertheless remain confined to the Ge-rich side, shifting only the composition at which \hmn{R3m} vanishes (\autoref{sec:E}).
The same heating protocol was applied to all investigated compositions spanning the full composition range for constructing the composition--temperature phase diagram presented in the following section.

\subsection{Composition--temperature phase diagram}\label{sec:E}

\begin{figure}[t]
    \includegraphics{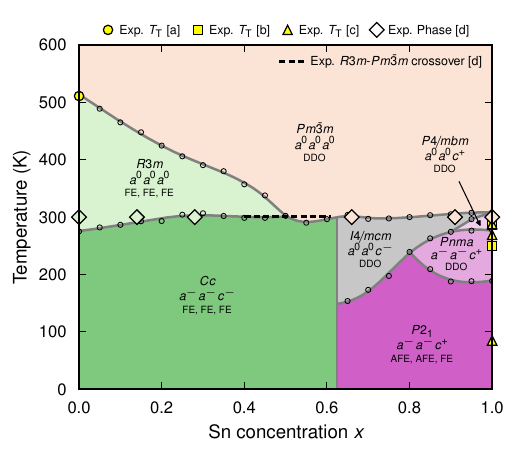}
\caption{
Composition--temperature phase diagram of \ce{CsGe_{1-$x$}Sn_{$x$}Br3} obtained from \gls{md} simulations using the \gls{qnep}.
Colored regions denote the stable structural phases.
The corresponding octahedral tilt systems (Glazer notation) and ordering of the $B$-site off-centering (FE, ferroelectric; AFE, antiferroelectric; DDO, dynamically disordered off-centering) are indicated for each phase.
Small black circles mark the transition temperatures determined from the heating runs, and the gray curves are guides to the eye representing the calculated phase boundaries.
Additional symbols denote available experimental transition temperatures ($T_\mathrm{T}$) and phase assignments from Refs. a \cite{thiele1987kristallstrukturen}, b \cite{fabini2016dynamic}, c \cite{fabini2024noncollinear}, and d \cite{hooper2026synthetic}.
The dashed line indicates the experimentally reported \hmn{R3m}--\hmn{Pm-3m} crossover.
}
    \label{fig:phase-diagram}
\end{figure}

The composition--temperature phase diagram assembled from these heating runs is shown in \autoref{fig:phase-diagram}.
Boundaries were determined for compositions spanning the full range in intervals of $\Delta x = 0.05$.
At low temperature the diagram is divided by the vertical \gls{mpb} between $x = 0.60$ and $0.65$ established in \autoref{sec:C}, separating the ferroelectric monoclinic \hmn{Cc} phase from the \hmn{P2_1} phase with mixed antipolar and polar off-centering.

The evolution with Sn content follows the energetic balance mapped in \autoref{sec:B}.
\ce{CsGeBr3} possesses a strong ferroelectric instability, whereas \ce{CsSnBr3} has weaker polar distortions and a stronger tendency toward octahedral tilting.
Increasing the Sn concentration therefore suppresses the magnitude and stability of the ferroelectric off-centering while favoring octahedral rotations, and this competition drives the transition from the \hmn{Cc} ground state to the \hmn{P2_1} phase at the boundary.

The finite-temperature behavior changes systematically across the composition range.
On the Ge-rich side the \hmn{Cc} phase transforms into the rhombohedral \hmn{R3m} phase before reaching the dynamically disordered cubic phase; the stability range of \hmn{R3m} narrows continuously with increasing Sn content and vanishes around $x \approx 0.5$, before the boundary is reached, consistent with the progressive suppression of long-range ferroelectric order.
The Sn-rich side is more composition dependent.
Immediately beyond the boundary the \hmn{P2_1} phase transforms on heating into an intermediate tetragonal \hmn{I4/mcm} phase before reaching the cubic phase; at larger Sn content the orthorhombic \hmn{Pnma} phase becomes stable, and near the Sn end member the sequence approaches that of \ce{CsSnBr3}, \hmn{P2_1} $\rightarrow$ \hmn{Pnma} $\rightarrow$ \hmn{P4/mbm} $\rightarrow$ \hmn{Pm-3m}.
For $0.80 \leq x \leq 0.95$ the \hmn{Pnma} phase transforms into \hmn{I4/mcm} rather than \hmn{P4/mbm} on heating, so that the in-phase tilt vanishes before the out-of-phase tilts, opposite to the sequence in \ce{CsSnBr3}.

The emergence of \hmn{I4/mcm} near the boundary follows from the competing tilt patterns of the neighboring low-temperature phases.
The Ge-rich \hmn{Cc} phase contains only out-of-phase rotations ($a^-a^-c^-$), whereas the Sn-rich \hmn{P2_1} phase combines out-of-phase rotations with an in-phase component ($a^-a^-c^+$).
As temperature suppresses the long-range off-centering and reduces the rotational amplitudes, the mixed compositions retain only a single out-of-phase tilt component, producing the $a^0a^0c^-$ pattern of \hmn{I4/mcm}.
This phase is therefore an intermediate rotational state between the all out-of-phase pattern favored on the Ge-rich side and the mixed in-phase/out-of-phase pattern stabilized at higher Sn content, a structure that exists only because the two parent tilt systems are incompatible, and one that has no counterpart in either end member.

The available experimental data are included in \autoref{fig:phase-diagram}.
For the end members, the transition temperatures of \ce{CsGeBr3} \cite{thiele1987kristallstrukturen} and \ce{CsSnBr3} \cite{fabini2016dynamic, fabini2024noncollinear} are reproduced as discussed in \autoref{sec:D}.
For the mixed compositions, Hooper \emph{et al.} assigned room-temperature structures from X-ray diffraction and found \hmn{R3m} for Ge-rich and \hmn{Pm-3m} for Sn-rich compositions, with a crossover between $x \approx 0.4$ and $0.6$ \cite{hooper2026synthetic}, in agreement with our phase diagram.
Kama \emph{et al.} reported the cubic structure at room temperature for spray-deposited films down to $x \approx 0.23$ \cite{kama2022sngebr3}, which places the crossover at lower Sn content.
The two experiments used different
synthesis routes, which are known to affect the structures obtained for these alloys \cite{hooper2026synthetic}.
Since the phase boundaries near room temperature are almost flat, small shifts in transition temperature correspond to large shifts in composition, and we consider the qualitative agreement with both experiments good.
The simulations further predict that the competition between stereochemically driven ferroelectric off-centering and octahedral tilting gives rise to an intermediate phase near the \gls{mpb} that is absent in the parent compounds.

\subsection{Band gap variations}

\begin{figure}[h]
    \includegraphics{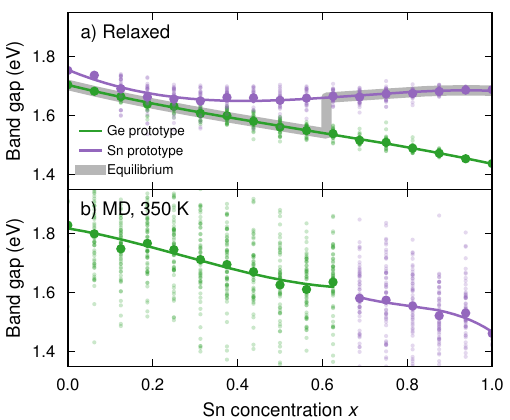}
\caption{
Band gap of \ce{CsGe_{1-$x$}Sn_{$x$}Br3} from PBEsol \gls{dft} calculations.
(a) Relaxed 80-atom structures started from the Ge (\hmn{Cc}) and Sn (\hmn{P2_1}) ground-state prototypes, with up to 30 random Sn/Ge occupations per composition.
Small points are individual occupations, large points their mean, and lines cubic fits to the means.
The gray line is the gap of the lower-energy prototype on each side of the crossover of the \gls{qnep} mixing energies at $x = 0.61$.
(b) 160-atom snapshots from \gls{md} simulations at \qty{350}{\kelvin} with the \gls{qnep}, started from the Ge prototype for $x \leq 0.625$ and the Sn prototype above.
Large points are means over 60 snapshots per composition (6 at the end members), lines cubic fits to the means.
}
    \label{fig:bandgap}
\end{figure}

Both octahedral tilting and $B$-site off-centering change the overlap between the $B$-site and halide orbitals that form the band edges \cite{filip2014steric, prasanna2017bandgap}, so the structural changes identified above should be reflected in the band gap.
\autoref{fig:bandgap}a shows the PBEsol band gap of the relaxed structures for both ground-state prototypes.
For the Ge prototype the gap decreases almost linearly from about \qty{1.7}{\electronvolt} at $x=0$ to \qty{1.4}{\electronvolt} at $x=1$, with only weak bowing, whereas for the Sn prototype it stays nearly constant at about \qty{1.7}{\electronvolt}.
Following the lower-energy prototype across the \gls{mpb}, the equilibrium gap jumps by about \qty{0.1}{\electronvolt} between $x=0.60$ and $0.65$, so the structural transition leaves a direct signature in the electronic structure.

\autoref{fig:bandgap}b shows the band gaps at \qty{350}{\kelvin}.
We chose this temperature instead of room temperature because several compositions undergo phase transitions close to \qty{300}{\kelvin} (\autoref{fig:phase-diagram}), and small errors in the predicted transition temperatures could then change which phase is simulated.
At \qty{350}{\kelvin} the picture changes.
On the Ge-rich side the gap is higher than for the relaxed structures by about \qty{0.1}{\electronvolt}, whereas on the Sn-rich side it is lower by up to \qty{0.2}{\electronvolt}.
As a result the jump at the \gls{mpb} disappears and the band gap decreases smoothly with Sn content over the whole composition range.
The opposite temperature dependence of the two sides reflects their different structures at \qty{350}{\kelvin}. The Ge-rich compositions retain strong local off-centering, whereas the Sn-rich compositions have lost their long-range tilting (\autoref{fig:phase-diagram}).

The decrease in band gap with increasing Sn content is consistent with the experimental trend reported for \ce{CsGe_{1-$x$}Sn_{$x$}Br3} thin films, where the optical gap decreases from approximately \qty{2.5}{\electronvolt} for \ce{CsGeBr3} to \qty{1.8}{\electronvolt} for \ce{CsSnBr3} \cite{kama2022sngebr3}.
While our PBEsol calculations underestimate the absolute values of the band gaps, we expect them to capture the qualitative trends in structure-induced band-gap changes, including the discontinuity at the \gls{mpb} and the temperature dependence.

\section{Conclusions}
We have used large-scale \gls{md} simulations with a charge-aware \gls{mlip} trained on hybrid-functional data to determine the composition--temperature phase diagram of \ce{CsGe_{1-$x$}Sn_{$x$}Br3}.
The alloy is not a simple interpolation between its end members.
The energy surfaces of \ce{CsGeBr3} and \ce{CsSnBr3} show that tilting and off-centering reinforce each other in both compounds, whereas the polar instability weakens and the tilts and their coupling to the off-centering strengthen from Ge to Sn.
Across the alloy series, this shift in balance leads to a first-order \gls{mpb} between $x = 0.60$ and $0.65$, which separates a ferroelectric \hmn{Cc} phase from a \hmn{P2_1} phase with mixed antipolar and polar off-centering, and, on heating, to an intermediate \hmn{I4/mcm} phase near the boundary that neither end member adopts.
The phase diagram reproduces the known transition sequences and temperatures of \ce{CsGeBr3} and \ce{CsSnBr3} and agrees with room-temperature diffraction on the alloys.
Above the transition temperatures the $B$ cations remain locally off-centered but orientationally disordered, so that the local structure of the cubic phase is polar even though its average structure is not.
The structural transition at the boundary is also reflected in the calculated band gap, which jumps at the boundary at low temperature but decreases smoothly with Sn content at \qty{350}{\kelvin}.
These results show that $B$-site mixing provides a direct handle on polar and rotational order in lead-free halide perovskites and identify the composition range near the boundary as the most promising for tuning their functional properties.

\section{Data Availability Statement}
The \gls{dft} data and \gls{qnep} model generated in this study are openly available via Zenodo at \url{https://doi.org/10.5281/zenodo.23955747}.

\section*{Acknowledgements}
We acknowledge funding from the Swedish Foundation for Strategic Research through a Future Research Leader programme (FFL21-0129), the European Research Council (ERC Starting Grant no. 101162195), and the Knut and Alice Wallenberg Foundation (Nos.~2023.0032 and 2024.0042).
This research project was financially supported by Mahasarakham University, Thailand.
The computations were enabled by resources provided by the National Academic Infrastructure for Supercomputing in Sweden (NAISS), partially funded by the Swedish Research Council through grant agreement no. 2022-06725, and by the Berzelius resource provided by the Knut and Alice Wallenberg Foundation at the National Supercomputer Centre (NSC).

The authors used AI tools (Claude, Anthropic, ChatGPT, and OpenAI) during the preparation of this manuscript for assistance with writing and language improvements.
All scientific content, interpretation, and final wording were reviewed and approved by the authors.

\bibliography{references}

\end{document}


\begin{center}

{\large %
Supporting Information
}
\vspace{2ex}

{\Large\bf %
Competing Octahedral Tilting and $B$-Site Off-Centering Govern the Phase Diagram of \ce{CsGe_{1-$x$}Sn_{$x$}Br3} Perovskite Alloys
}

\vspace{3ex}

Apinya Ngoipala,$^{1,2,*}$
Erik Fransson,$^2$
Paul Erhart,$^2$
and Julia Wiktor$^{2,\dagger}$

\end{center}

\begin{flushleft}

\newcommand{\addmsu}{Nanomaterials Physics Research Unit, Department of Physics, Faculty of Science, Mahasarakham University, Maha Sarakham 44150, Thailand}
\newcommand{\addchalmers}{Department of Physics and Astronomy, Chalmers University of Technology, SE-412 96, Gothenburg, Sweden}

$^1$ \addmsu{} \\
$^2$ \addchalmers{} \\
$^*$ apinya.n@msu.ac.th \\
$^\dagger$ julia.wiktor@chalmers.se
\end{flushleft}


\tableofcontents

\clearpage




\section{Supplementary Figures}

\begin{figure*}[ht]
    \centering
    \includegraphics{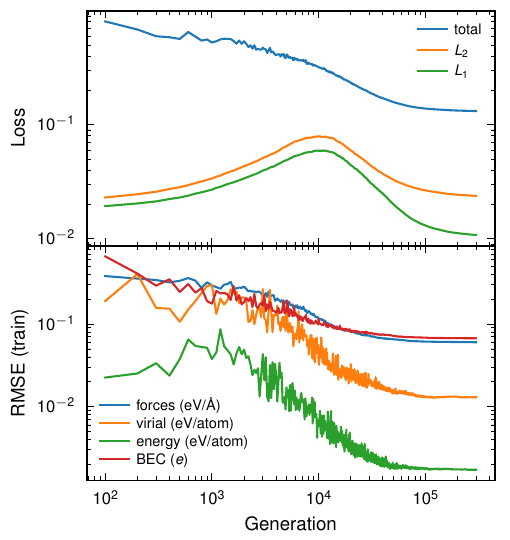}
    \titlecaptionlabel{
        Loss curves
    }{
        Evolution of the total loss, the $L_1$ and $L_2$ regularization terms, and the \glspl{rmse} for energies, forces, virials, and \glspl{bec} during training of the final \gls{qnep} model.
        The total loss is the sum of the weighted \glspl{rmse} and the two regularization terms.
    }{sfig:loss_curves}
\end{figure*}

\begin{figure*}[ht]
    \centering
    \includegraphics[width=\textwidth]{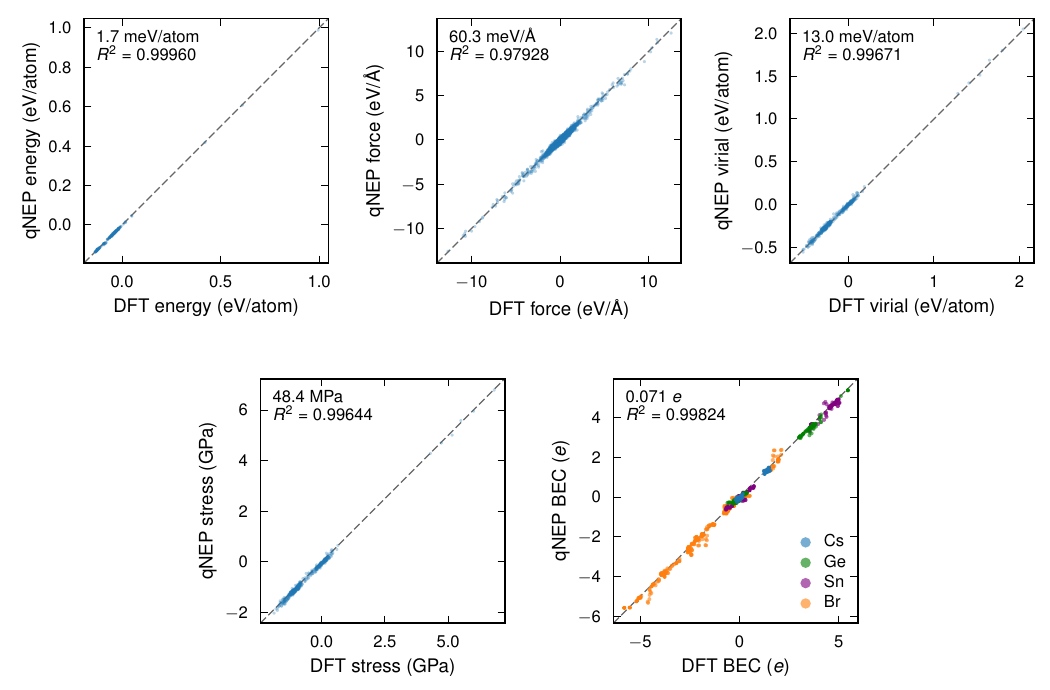}
    \titlecaptionlabel{
        Parity plots
    }{
        Parity plots for energies, forces, virials, stresses, and \gls{bec} tensor components predicted by the final \gls{qnep} model compared to the \gls{dft} reference data.
        The \glspl{bec} are colored by atomic species.
        Structures for which the stress tensors were not evaluated with \gls{dft} are not included in the virial and stress parity plots.
    }{sfig:parity_all}
\end{figure*}

\begin{figure*}[ht]
    \centering
    \includegraphics[width=\textwidth]{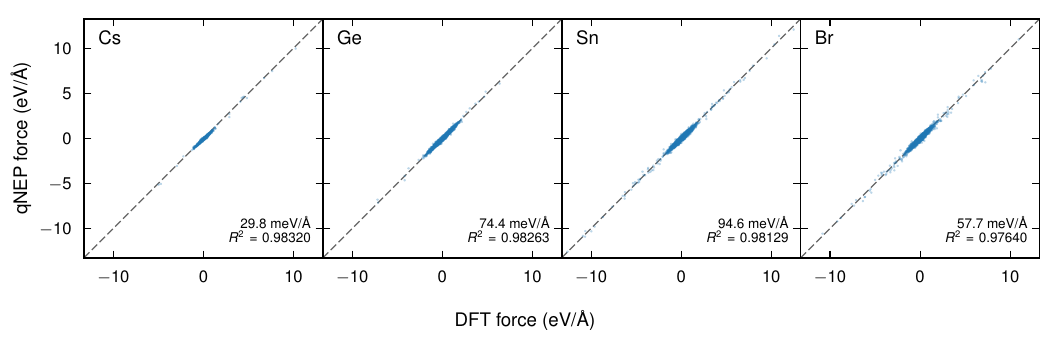}
    \titlecaptionlabel{
        Species-resolved force parity plots
    }{
        Force parity plots of the final \gls{qnep} model for each atomic species (Cs, Ge, Sn, and Br).
    }{sfig:parity_species}
\end{figure*}

\begin{figure*}[ht]
    \centering
    \includegraphics[scale=0.9]{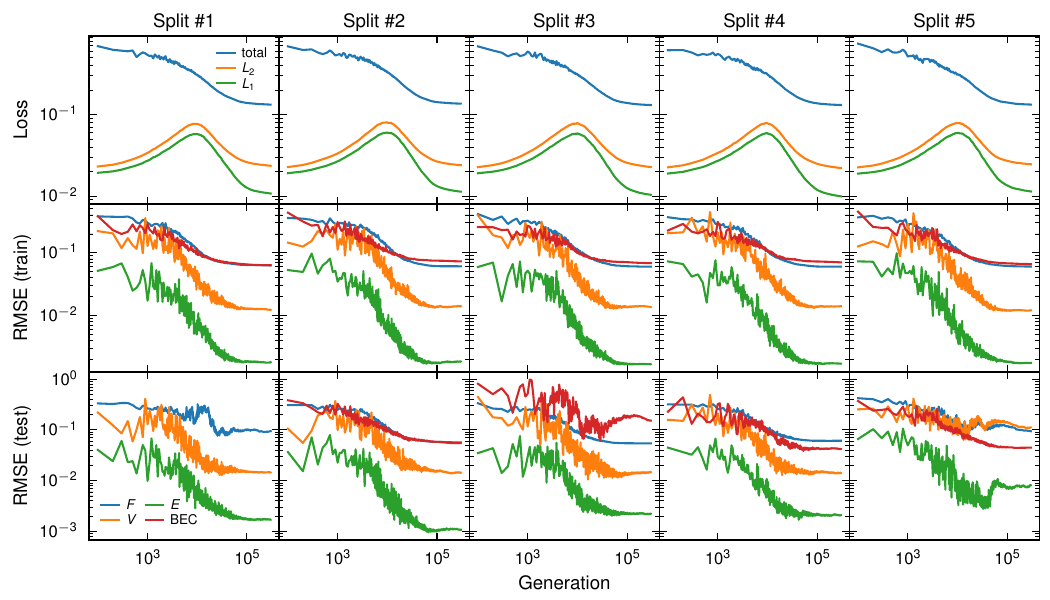}
    \titlecaptionlabel{
        Cross-validation loss curves
    }{
        Loss curves for the five models trained in the five-fold cross-validation.
        For each split, the panels show the total loss and the $L_1$ and $L_2$ regularization terms (top) and the \glspl{rmse} for forces (F), virials (V), and energies (E) on the training set (middle) and on the test set (bottom).
    }{sfig:loss_curves_splits}
\end{figure*}

\clearpage
\section{Supplementary Tables}

\begin{table}[H]
\centering
\small
\titlecaptionlabel{
    Training and cross-validation errors
}{
    \Glspl{rmse} and coefficients of determination ($R^2$) of the \gls{qnep} model.
    ``Full'' refers to the final model trained on the complete data set (training errors).
    ``CV'' gives the mean $\pm$ standard deviation over the five cross-validation splits (\autoref{stab:qnep_errors_splits}).
}{stab:qnep_errors}
\begin{tabular}{llcccccc}
\toprule
 & & \multicolumn{2}{c}{Full (train)} & \multicolumn{2}{c}{CV train} & \multicolumn{2}{c}{CV test} \\
\cmidrule(lr){3-4}\cmidrule(lr){5-6}\cmidrule(lr){7-8}
Property & Unit & RMSE & $R^2$ & RMSE & $R^2$ & RMSE & $R^2$ \\
\midrule
Energy & meV/atom & 1.7 & 0.9996 & 1.8 $\pm$ 0.1 & 0.9996 $\pm$ 0.0001 & 3.1 $\pm$ 2.9 & 0.9977 $\pm$ 0.0015 \\
Force & meV/\AA & 60.3 & 0.9793 & 61.3 $\pm$ 1.2 & 0.9790 $\pm$ 0.0009 & 72.3 $\pm$ 20.5 & 0.9626 $\pm$ 0.0180 \\
Virial & meV/atom & 13.0 & 0.9967 & 13.3 $\pm$ 1.0 & 0.9964 $\pm$ 0.0005 & 33.9 $\pm$ 43.4 & 0.9748 $\pm$ 0.0333 \\
Stress & MPa & 48.4 & 0.9964 & 49.8 $\pm$ 3.5 & 0.9961 $\pm$ 0.0005 & 117.6 $\pm$ 142.2 & 0.9763 $\pm$ 0.0316 \\
BEC & $e$ & 0.071 & 0.9982 & 0.073 $\pm$ 0.004 & 0.9982 $\pm$ 0.0002 & 0.074 $\pm$ 0.053 & 0.9978 $\pm$ 0.0025 \\
\bottomrule
\end{tabular}
\par\smallskip\footnotesize The BEC CV-test values are averaged over splits 2--5, since the test set of split 1 contains no BEC data.
\end{table}

\begin{table}[H]
\centering
\small
\titlecaptionlabel{
    Errors for each cross-validation split
}{
    \Glspl{rmse} of the \gls{qnep} model ($R^2$ in parentheses) on the training and test sets of each of the five cross-validation splits.
    Each model was trained on 313 structures and tested on the remaining 35.
}{stab:qnep_errors_splits}
\begin{tabular}{llccccc}
\toprule
Split & Set & Energy (meV/atom) & Force (meV/\AA) & Virial (meV/atom) & Stress (MPa) & BEC ($e$) \\
\midrule
1 & Train & 1.8 (0.9996) & 63.4 (0.9777) & 12.4 (0.9972) & 45.9 (0.9970) & 0.067 (0.9984) \\
 & Test & 1.7 (0.9985) & 94.9 (0.9323) & 14.6 (0.9883) & 53.6 (0.9895) & -- \\
\addlinespace
2 & Train & 1.9 (0.9996) & 61.2 (0.9795) & 14.2 (0.9963) & 53.2 (0.9959) & 0.079 (0.9979) \\
 & Test & 1.1 (0.9995) & 56.8 (0.9740) & 14.3 (0.9887) & 53.1 (0.9893) & 0.055 (0.9986) \\
\addlinespace
3 & Train & 1.7 (0.9996) & 60.6 (0.9799) & 13.8 (0.9964) & 51.3 (0.9961) & 0.072 (0.9981) \\
 & Test & 2.2 (0.9971) & 54.3 (0.9745) & 14.7 (0.9896) & 55.9 (0.9904) & 0.153 (0.9940) \\
\addlinespace
4 & Train & 1.7 (0.9996) & 60.6 (0.9796) & 14.2 (0.9963) & 52.5 (0.9960) & 0.074 (0.9980) \\
 & Test & 2.1 (0.9980) & 61.2 (0.9729) & 14.2 (0.9922) & 53.4 (0.9925) & 0.042 (0.9995) \\
\addlinespace
5 & Train & 1.8 (0.9995) & 60.6 (0.9785) & 12.2 (0.9959) & 46.2 (0.9956) & 0.071 (0.9983) \\
 & Test & 8.1 (0.9956) & 94.4 (0.9595) & 111.6 (0.9154) & 371.9 (0.9197) & 0.045 (0.9991) \\
\bottomrule
\end{tabular}
\par\smallskip\footnotesize
The test set of split 1 contains no BEC data.
The larger energy, virial, and stress test errors of split 5 all come from a single high-energy configuration, lying about \qty{0.6}{\electronvolt\per\atom} above the rest of the test set and with a virial several times larger than that of any other test structure.
Without this structure, the split-5 test \glspl{rmse} are \qty{2.4}{\milli\electronvolt\per\atom} for the energy, \qty{16.6}{\milli\electronvolt\per\atom} for the virial, and \qty{61.0}{\mega\pascal} for the stress.
This structure is included in the training set of the final model.
\end{table}
